\documentclass[prb,twocolumn,showpacs]{revtex4}
\usepackage{graphicx}
\usepackage{dcolumn}
\usepackage{bm}

\begin{document}

\title{Wavepacket spreading dynamics  under a non-instantaneous nonlinearity: Self-trapping, defocusing and focusing}

\author{Marcelo L. Lyra and Rodrigo P.\ A.\ Lima}
\affiliation{Instituto de F\'{\i}sica, Universidade Federal de Alagoas 
57072-970 Macei\'o-AL, Brazil}

\date{\today}

\begin{abstract}

Special localized wavemodes show up in several physical scenarios including BEC in optical lattices, nonlinear photonic crystals and  systems with strong electron-phonon interaction. These result from an underlying nonlinear contribution to the wave equation that is usually assumed to be instantaneous. Here we demonstrate that the relaxation process of the nonlinearity has a profound impact in the wavepacket dynamics and in the formation of localized modes. We illustrate this phenomenology by considering the one-electron wavepacket spreading in a $C60$ buckball structure whose dynamics is governed by a discrete nonlinear Schroedinger equation with a Debye relaxation of the nonlinearity. 
We report the full phase-diagram related to the spacial extension of the asymptotic wavepacket and unveil a complex wavepacket dynamical behavior. 

\end{abstract}

\pacs{71.30.+h, 73.20.Mf, 05.45.-a, 73.61.Wp}
\maketitle

The discrete nonlinear Schroedinger equation (DNSE) has been frequently used to describe a wealthy of nonlinear physical phenomena associated with wave dynamics. The DNSE can display a large class of topologically stable solutions such as solitons, vortex rings and breathers\cite{braun,hennig,bishop,arevalo,law,bobec}. These solutions have been largely explored in the context of Bose-Einstein condensates in optical lattices and light propagation in photonic crystals\cite{bobecreview,wang,xue,ponomarenko}. The DNSE also captures some features related to the electron-phonon interaction that can be effectively taken into account by a nonlinear term in the electronic Hamiltonian\cite{molina}. Nonlinearity was shown to breakdown the Anderson localization in low-dimensional disordered systems by promoting a diffusive-like spread of the electronic wavepacket\cite{pikovsky}.

The self-trapping of the wavefunction is one of the most remarkable effects associated with the presence of a nonlinear contribution to the discrete Schroedinger equation. An initially localized wavepacket spreads in the regime of weak nonlinearities, but becomes trapped around its initial position when the nonlinear coupling is above a threshold value\cite{johansson,datta,pan}. The self-trapping transition has been characterized in one-dimensional and two-dimensional systems. The critical nonlinear coupling is of the order of the energy bandwidth of the corresponding linear system for an initial wavepacket fully localized in a single site. In two-dimensional lattices  it increases as the initial wavepacket becomes wider\cite{dias}.  Recently, the influence of the nonlinear response time on the self-trapping transition has been investigated in one-dimensional systems. By considering a delayed nonlinearity, it has been shown that the critical nonlinearity depends non-monotonically on the response time\cite{demoura}. When the relaxation of the nonlinearity is described by a Debye process, the typical response time drastically affects the wavepacket dynamics. Much weaker nonlinearities are needed to trap the wavepacket\cite{vidal}. It focalizes after an initial spread exhibiting pronounced finite-size effects including  wavepacket fragmentation. The relaxation of the nonlinearity also leads to the wavepacket relocalization in disordered systems\cite{caetano,pikovsky2}.

The wavepacket dynamics in nonlinear honeycomb and some closely related lattices has been a subject of intense recent investigations. These have been mainly motivated by the ability to build up optical lattices and photonic crystals with these topologies, which allows the experimental observation of several theoretically predicted phenomena, such as topology-induced bistability\cite{molina3}, discrete localized modes and breathers\cite{law5,leblond}, conical diffraction\cite{zhu,peleg2}, self-trapping of vortices,  gap vortices and gap solitons\cite{law4}. Nonlinearity is also expected to play a significant role in the dynamics of electronic wavepackets in carbon-based structures\cite{hartmann,brihaye}. Recent experiments have probed the electron-phonon interaction in graphene which can accounts for an effective nonlinearity\cite{bianchi}.  Low-temperature scanning tunneling spectroscopy experiments have been recently used to map the wavefunction in graphene quantum dots\cite{libisch}, which open the possibility to experimentally observe many theoretically predicted nonlinear effects in low-dimensional carbon-based structures. However, most of the previous theoretical  studies of nonlinear wave dynamics in two-dimensional lattices do not include finite-size effects and the relaxation of the nonlinearity  which might play important roles in nano-scaled structures.

In the present work, we show that both finite-size and finite response-time effects have a strong impact on the wavepacket spreading dynamics. We will follow the time-evolution of an initially fully localized wavepacket whose propagation is governed by a DNSE with Debye relaxation of the nonlinearity. To illustrate the phenomenology, we consider the wavepacket evolution over a $C60$ buckball. In this topology, all sites are equivalent and the absence of borders avoids the superposition of surface effects to the finite-size ones. We will report the phase diagram in the two dimensional parameter space (nonlinear coupling $\chi$ and response time $\tau$) which unfolds a quite complex structure with three distinct localized phases.

A nonlinear contribution for the one-electron Schroedinger equation results from the underlying electron-phonon interaction\cite{molina,pikovsky,johansson,datta,pan,dias}. Within the adiabatic approximation, such nonlinear term can be considered as instantaneous, with the on-site potential at a given time depending on the local electronic density at this proper time. Going beyond the adiabatic approximation, the effective nonlinear term has been shown to obey a Debye relaxation process\cite{kenkre1}. Within a tight-binding approach and considering a localized orbitals basis, the dynamics of a one-electron wavepacket can be described by a DNSE with a relaxing nonlinearity, written as
\begin{eqnarray}
i\dot{\Psi}_n(t) &=& \varepsilon_n\Psi_n(t) +\sum_m V_{nm}\Psi_m(t) - X_n(t)\Psi_n(t) \\
\dot{X}_n(t)&=&-\frac{1}{\tau}\left[X_n(t)+\chi|\Psi_n(t)|^2\right] ~,
 \end{eqnarray}
where we used $\hbar=1$. $\Psi_n(t)$ is coefficient
of the wave vector expanded in the localized orbitals basis $|\Psi(t)\rangle = \sum_n\Psi_n(t)|n\rangle$. The relaxing nonlinearity $X(t)$ is considered to have a typical response time $\tau$ and strength coefficient $\chi$. The sum is taken over first-neighbor pairs of sites $(n,m)$ that are coupled by a hopping amplitude $V_{nm}$.   $\varepsilon_n$'s represent the on-site  energies. In the limit of $\tau\rightarrow 0$, the nonlinearity becomes instantaneous $X_n(t) = -\chi|\Psi_n(t)|^2$ and the usual DNSE is recovered. 

In what follows, we will solve the above nonlinear dynamical equations for an initially localized wavepacket $\Psi_n(t=0) =\delta_{n,1}$. We will consider a $C60$ buckball topology (see fig.~1) which is composed of 20 hexagons and 12 pentagons. Each of the $60$ equivalent sites have three bonds. Two of them are between a hexagon and a pentagon and the remaining one between two hexagons. In the real $C60$ molecule, the hexagon-hexagon bonds are slightly shorter than the hexagon-pentagon bonds. This feature ultimately leads to a small difference in the corresponding hopping amplitudes which we will not consider in the present work. We will take $V_{nm}=V$ for every bond and will work in units of $V=1$. Further, as all sites are equivalent, we can also consider $\varepsilon_n=\varepsilon=0$ without any loss of generality. The dynamical equations were solved using a standard eighth-order Range-Kutta algorithm and the norm conservation was followed with a precision of $10^{-8}$ to ensure the numerical accuracy.

In order to characterize the spacial extension of the wavepacket, we computed the participation function defined as 
\begin{equation}
 P(t) = \left[\sum_{n=1}^N|\Psi_n(t)|^4\right]^{-1} ~,
\end{equation}
which gives a measure of the fraction of the lattice sites at which the wavepacket is spread at time $t$. It becomes equal to $N$ for an uniformly distributed wavepacket. For a fully localized state $P=1$.

We followed the time-evolution of the wavepacket until it reached a statistically stationary regime. We explored the two-dimensional parameter space $(\tau,\chi )$ in order to build up the full phase-diagram regarding the spacial extension of the asymptotic wavepacket. Four distinct phases were identified, as reported in figure 2. There is a single extended phase, on which the wavepacket spreads uniformly over all sites of the buckball, and three distinct localized phases. The extended phase prevails in the regime of weak nonlinearities. However, narrow branches of the extended phase are found between localized phases. While these extended branches die out when the relaxation time increases, the main extended phase becomes wider. 

The predominant localized phase $L_1$ corresponds to an asymptotic wavepacket that results localized mostly at the initial position (site labelled $C_1$ in fig.~1). In this region, we found the participation number to become slightly above unit, signalling a quite localized state, with the probability density concentrated at the initial position. There are two main regions of the parameter space on which the wavepacket becomes trapped on the initial position. A second localized phase $L_2$ appears in between these two $L_1$ regions. In this new localized phase, the participation function is of the same order as in the localized phase $L_1$. However, the probability density is centered at the site that is diametrically opposite to the initial position (labeled as $C_2$ in fig.~1).  The two branches of the extended phase are in between these two localized phases. A third localized phase $L_3$ occurs in between $L_1$ and $L_2$. In this narrow phase, the wavepacket was found to become localized around other positions besides the initial and the diametrically opposite sites. It is interesting to call attention to the fact that a complex sequence of localization-delocalization transitions takes place at small values of the relaxation time when one crosses the regime of intermediate nonlinear strengths.

\begin{figure}
\begin{center}
\includegraphics[width=60mm,clip]{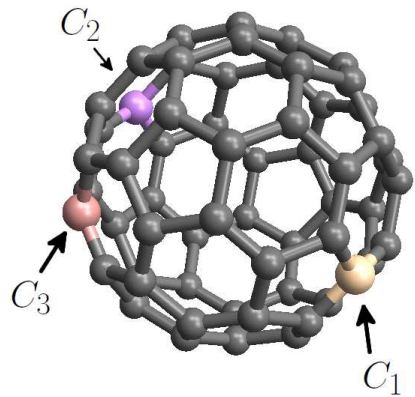}
\caption{(Color online) Schematic representation of the $C60$ buckball. Each site has two hexagon-hexagon bonds and one hexagon-pentagon bond. $C_1$ represents the site at which the wavepacket is fully localized at $t=0$. $C_2$ is the site diametrically opposite to the initial position. $C_3$ stands for the site targeted in figure 3d.} 
\label{fig1}
\end{center}
\end{figure}

\begin{figure}[ht!]
\centerline{\includegraphics[width=85mm,clip]{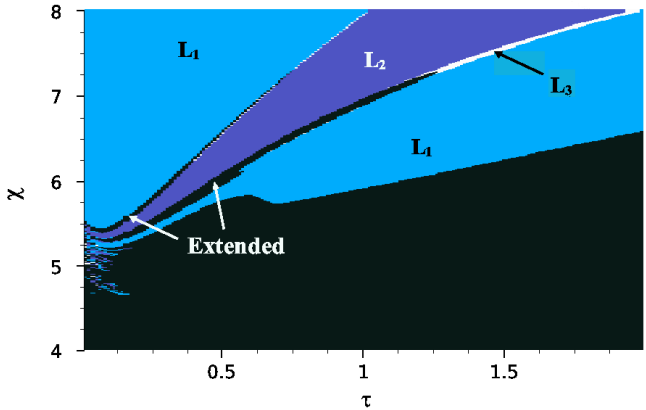}}
\caption{(Color online) Phase diagram concerning the spacial extension of the asymptotic wavepacket in the ($\tau , \chi$) parameter state. The extended phase with a uniformly distributed wavepacket occurs at small nonlinearities and two branches. $L_1$ corresponds to a phase with the wavepacket localized around its initial position. $L_2$ is the phase where the localization is concenterated in the site diametrically opposite to the initial position. In the narrow $L_3$ phase, the wavepacket localizes in other sites. A complex sequence of localization-delocalization transitions is observed at small $\tau$ values and intermediate nonlinear strengths.} 
\label{fig2}
\end{figure}

In order to have a clearer view of the wavepacket dynamics on these distinct phases, we plot the time-evolution of the participation function and the relevant occupation probabilities for some representative values of the physical parameters, as shown in fig.~3. The dynamics in the extended phase is illustrated in fig~3a for three set of parameter values which cover the two branches and the main extended region. The normalized participation function converges to $P/N=1$ indicating that the asymptotic wavepacket is uniformly distributed over all buckball sites. The convergence to this state is achieved after an oscillatory transient regime, thus indicating that the expanding wavepacket has a breathing character.

In fig.~3b we depict the typical wavepacket dynamics in the $L_1$ localized phase. We show the time-evolution of both the participation number and the occupancy probability of the initial site $|\Psi_1(t)|^2$. In the initial transient, the wavepacket spreads in an oscillatory fashion and occupies a finite fraction of the buckball sites. After this transient, the wavepacket self-focuses and relocalizes. The increase of $|\Psi_1(t)|^2$ indicates that the wavepacket relocalizes around its initial position. This relocalized state was found to be stable over a very long run.  A very similar phenomenon occurs in the localized phases $L_2$ and $L_3$, as illustrated in fig.3c and fig.~3d respectively. 
However, the self-focusing drives the wavepacket towards the site diametrically opposite to its initial position in phase $L_2$. In the localized $L_3$ phase, the wavepacket spreads almost uniformly over the lattice before starting to self-focuses. In this case, the position of the asymptotic localized state is very sensitive to values of the physical parameters. For the values used in fig.~3d, the wavepacket relocalizes around the site labeled as $C_3$ in fig.~1. 

\begin{figure}[ht!]
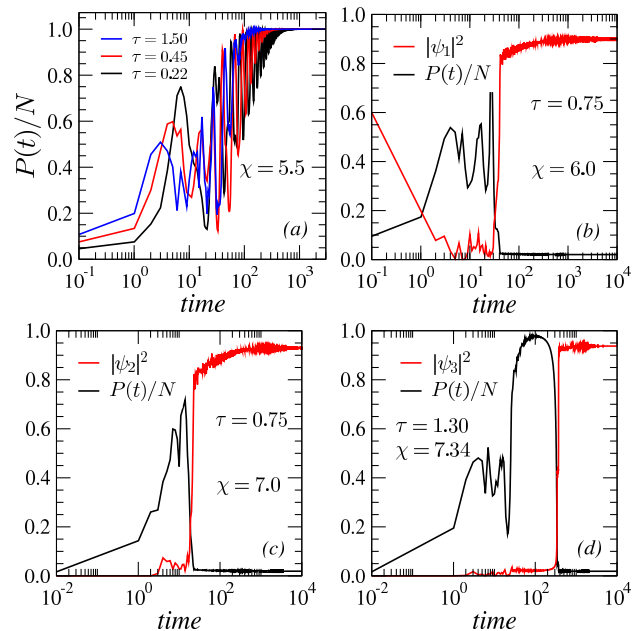

\centering
\begin{tabular}{cc}
\includegraphics[width=42mm,clip]{fig3a.eps} &
\includegraphics[width=39mm,clip]{fig3b.eps} \\
\includegraphics[width=39mm,clip]{fig3c.eps} &
\includegraphics[width=39mm,clip]{fig3d.eps}
\end{tabular}
\caption{(Colour online) Time-evolution of the normalized particpation number $P/N$ and the relevant occupancy probabilities $|\Psi_i(t)|^2$ for representative values of the physical parameters. (a) Extended phase: $\chi=5.5$ and $\tau = 0.22, 0.45$ and $1.50$ (from right to left) being respectively on the first, second and third extended regions. Notice that the expanding wavepacket develops breathing oscillations. (b) Localized $L_1$ phase: after the initial expansion, the wavepacket self-focuses and relocalizes around its initial position. (c) Localized $L_2$ phase: in this case, the wavepacket relocalization occurs around the site diametrically opposite to its initial position. Localized $L_3$ phase: here the relocalization occurs around the site labbelled as $C_3$ in fig.~1.} 
\label{fig3}
\end{figure}

\begin{figure}[ht!]
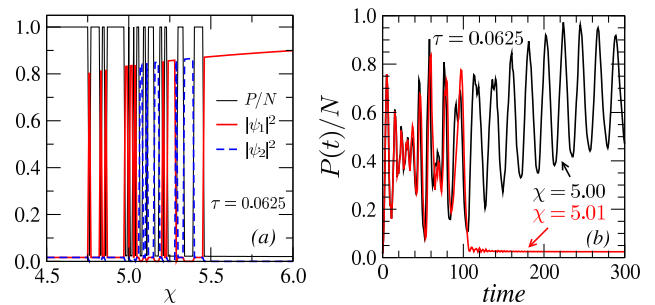

\centering
\begin{tabular}{cc}
 \includegraphics[width=39mm,clip]{fig4a.eps} &
\includegraphics[width=43mm,clip]{fig4b.eps}
\end{tabular}
\caption{(Color online) (a) Asymptotic normalized participation function $P/N$ an occupancy probabilities $|\Psi_i(t)|^2$ at the initial ($i=1$) and diametrically opposite ($i=2$) sites as a function of the nonlinear coupling for $\tau = 0.0625$. The system depicts a complex sequence of localization and delocalization transitions. (b) The time-evolution of the normalized participation function for $\tau=0.0625$ and two slightly different nonlinear strengths. Notice a dynamical transition that leads to distinct phases.} 
\label{fig4}
\end{figure}

Before finishing, we closely analyze the wavepacket dynamics in the parameter region where a complex sequence of localization-delocalization transitions occurs within the main extended region (see fig.~4). We firstly plot the asymptotic normalized participation function and occupancy probabilities of the initial and diametrically opposite sites as a function of the nonlinear coupling and a fixed relaxation time $\tau=0.0625$, as shown in fig.~4a.   $P/N=1$ in the extended phase, as expected for a uniformly delocalized wavepacket. However, before the ultimate localization transition that traps the wavepacket around its initial position, a complex sequence of localization-delocalization transitions takes place. The position of the localized wavepacket appears to randomly jump between the initial and diametrically opposite sites.  For other small values of the relaxation time, we verified that the wavepacket can eventually relocalizes around other sites. The high sensitivity of the wavepacket dynamics on the precise values of the parameters in this region is illustrated in fig.~4b. One observes that, while the orbits of two cases with slightly different parameters remain quite close to each other during some time, a dynamical transition takes them apart. It is worth to recall that a similar dynamical transition has been previously reported to occur in a two-level system driven by a relaxation process of the nonlinearity\cite{kenkre1}.

In conclusion, we showed that the wavepacket spreading dynamics displays a rich phenomenology including self-trapping, defocusing and focusing when the relaxation process of an effective third-order nonlinearity is taken into account. A $C60$ buckball structure was used to illustrate this phenomenology.  By considering a fully localized initial state, we provided the full phase diagram for the spacial extension of the long-time wavepacket. The phase diagram was shown to have a complex structure unveiling the presence of three distinct phases of localized wavepackets, besides the extended phase in which the wavepacket spreads uniformly over all buckball sites. The extended phase predominates at small nonlinearities. In the localized phases, the wavepacket was shown to relocalize after an initial spreading over a finite fraction of the network sites. Besides the usual relocalization of the wavepacket around its initial position, focalization on the diametrically opposite site takes place in a large stripe of the phase diagram. A narrow phase where the wavepacket relocalizes around other sites was also identified. A complex sequence of localization-delocalization transitions was shown to occur at small relaxation times as a function of the nonlinear coupling. These were associated to a dynamical transition which is strongly sensitive to the parameter values in this region. A stability analysis of the stationary states would be in order to shed light in the underlying mechanism that drives these transitions.

The main aspects of the here reported phenomenology are expected to show up in general nonlinear physical systems where the wavepacket dynamics is driven by a relaxation process of an underlying nonlinearity. These include electronic states in nano-sized clusters with strong electron-phonon coupling, BEC in optical lattices and light propagation in nonlinear photonic crystals.  
The present work leaves open the quest of investigating the possible influence of open boundaries and saturation effects that might also be relevant in optical and BEC experiments in finite planar lattices. 
With the recent advances of experimental techniques that allows to map a diversity of localized wave modes, further theoretical efforts would be in order to evaluate the relative relevance of these ingredients to the wavepacket dynamics in the presence of a relaxing nonlinearity.

We would like to thank CAPES via project PPCP-Mercosul,  CNPq, and FINEP (Brazilian
Research Agencies) as well as FAPEAL (Alagoas State
Research Agency) for partial financial support. We also would like to thank the kind hospitality of
the Facultad de Matem\'atica, Astronomia y F\'{\i}sica (FaMaF) of Universidad Nacional de Cordoba where part of this work was developed.

\end{document}